\documentclass[aps,prb,twocolumn,showpacs,superscriptaddress]{revtex4}%
\usepackage{epsfig,amsmath,amssymb}
\usepackage{graphicx}
\usepackage[cp1251]{inputenc}
\usepackage[english]{babel}

\begin{document}

\title{Coherent Rabi response of a charge-phase qubit under
microwave irradiation} %
\author{V.I. Shnyrkov}
\email{shnyrkov@ilt.kharkov.ua} %
\affiliation{B. Verkin Institute for Low Temperature Physics and
Engineering, National Academy of Sciences of Ukraine, av. Lenina 47,
Kharkov 61103, Ukraine}%
\author{D. Born}
\altaffiliation[D.B. is now with ] {Scuola Normale Superiore di
Pisa, NEST Piazza dei Cavalieri 7, 56126 Pisa, Italy.} %
\affiliation{Friedrich Schiller University, Institute of Solid State
Physics, Helmholtzweg 5, D-07743 Jena, Germany} %
\author{A.A. Soroka}
\affiliation{National Science Center "Kharkov Institute of Physics
and Technology", Akhiezer Institute for Theoretical Physics, st.
Akademicheskaya 1, 61108 Kharkov, Ukraine}
\author{W. Krech}
\affiliation{Friedrich Schiller University, Institute of Solid State
Physics, Helmholtzweg 5, D-07743 Jena, Germany} %

\begin{abstract}

We report on radio-frequency measurements of the charge-phase qubit
being under continuous microwave irradiation in the state of weak
coupling to a radio-frequency tank circuit. We studied the rf
impedance dependence on the two important parameters such as power
of microwave irradiation whose frequency is close to the gap between
the two lowest qubit energy levels, and temperature of the internal
heat bath. We have found that backaction effects of the qubit on the
rf tank, and vice versa, tank on the qubit, lead to a negative as
well as a positive real part of the qubit impedance Re$Z(\omega)$
seen by the tank. We have implemented noise spectroscopy
measurements for direct impedance readout at the extreme points
corresponding to maximum voltage response and obtained absolute
values of about $0.017\,\Omega$ for the negative and positive
Re$Z(\omega)$. Our results demonstrate the existence and persistence
of the coherent single- and multi-photon Rabi dynamics of the qubit
with both negative and positive dynamic resistance inserted into the
tank in the temperature range of 10 to 200 mK.
\end{abstract}

\pacs{03.75.Lm, 74.50.+r, 85.25.Cp}

\maketitle

\section{Introduction}

Nowadays the problem is under intensive discussion of building a new
class of information signal detectors based on solid-state Josephson
qubits in which coherent phenomena like single- and multi-photon
Rabi oscillations occur. \cite{R1,R2,R3,R4,R5,R6,R7} The behavior of
these detectors would essentially differ from that of quantum
devices utilizing non-linear effects of macroscopic quantum
tunnelling, macroscopic resonant tunnelling and superposition of
macroscopic states by the presence of intrinsic generation at the
Rabi frequency. The variation of ground and excited levels
population with the Rabi frequency results in a periodic alteration
of the magnitude and sign of reactive parameter of the detector.
Similar to the known classical electrodynamics of Josephson junction
in a resonator,\cite{Kanter,Likharev} the interaction of a weak
signal from the tank circuit with Rabi oscillations may lead to
synchronization effects and emerging the negative resistance, i.e.
the signal amplification.\cite{Smirnov1,Krech1,Greenberg} The
amplitude of these effects will depend on the qubit decoherence time
$\tau_\varphi$ (Rabi oscillations frequency band), the latter must
exceed the period of oscillations in the tank circuit coupled with
the qubit, $\tau_\varphi > T=2\pi/\omega_T,$ where $\omega_T$ is the
tank circuit resonant frequency.

It was shown\cite{Sisyphus1,Sisyphus2} that a qubit placed in
microwave field might show Sisyphus cooling
effects\cite{Sisyphus2,FQ,FQ1&2} for the opposite limit
$\tau_\varphi < T \approx \tau_R$, i.e. when the qubit decoherence
time was less than the $LC$ tank circuit oscillation period which
was about the relaxation time $\tau_R$. The observation of the
negative resistance inserted by the qubit into the tank circuit is
possible in both cases. However, the resistance value and its
dependence on temperature, frequency and microwave power will
substantially differ. In both cases, the characteristics of the
qubit-detector system will depend on the qubit's effective
temperature $T_{eff}$, which is contributed by the measurement
process itself, the physical temperature of the bath $T_{bath}$, the
quality of thermalization\cite{therm} of the qubit input circuit
(gates), etc. Note that, in a region of large $T_{eff}$ values, the
observation of complicated classical electrodynamics is possible
associated with so called pseudo-Rabi oscillations.\cite{P-Rabi}

The aim of this work is to experimentally analyze coherent processes
in a charge-phase qubit\cite{Zorin,Krech2} being under continuous
microwave irradiation in the state of weak coupling to a
radio-frequency tank circuit. It is important to note here that our
experiment is similar but not identical to that of Ref.
\onlinecite{Sisyphus2}. In this paper we address the problem of the
coherent single- and multi-photon Rabi dynamics of the charge-phase
qubit with $\tau_\varphi > T$ in a wide temperature range. The
direct observation of absorption and irradiation properties of a
strongly driven qubit coupled to a rf tank circuit, being of
fundamental interest, also has potential applications for quantum
detectors based on charge-flux qubits with self-Rabi frequency
pumping.

The paper is organized as follows. In Sec.\,II theoretical
description of a strongly driven charge-phase qubit coupled to an rf
tank-detector and analysis of measured output functions are given.
The main Sec. III is devoted to detailed experimental investigation
of coherent single- and multi-photon Rabi dynamics of the strongly
driven charge-phase qubit probed by rf tank detector.

\section{System of charge-phase qubit and tank circuit}

\begin{figure}[t!]
\centering %
\scalebox{1.05}[1.05]{\includegraphics[bb = 36 373 263 563]{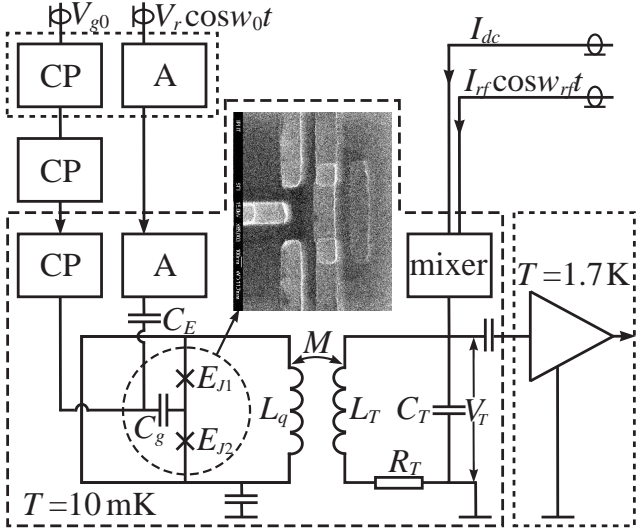}} %
\caption{\label{fig01} Circuit diagram of the charge-phase
measurement system. The components situated at the lowest
temperature are shown in the dashed box. The qubit and $LC$ tank
circuit parameters are: $L_q\simeq 0.9\,$nH, $I_{c1,2}\simeq
100\,$nA, $C_g\simeq 0.18\,$fF, $C_{J1,2}\simeq 2.0\,$fF, $L_T\simeq
0.17\,\mu$H, $C_T\simeq 165\,$pF, $\omega_T/2\pi=29.8\,$MHz,
$R_T\simeq 0.1\,\Omega$, $Q_0\approx 350$; $M\approx 0.365\,$nH and
the parameter $k^2Q_0\approx 0.3.$ The preamplifier and the part of
the tank circuit placed on 1K-pot are shown in the dotted boxes. All
the electric lines are equipped with copper powder filters (CP) and
attenuators (A). }
\end{figure}

The charge-phase qubit (Fig.\,\ref{fig01}) is topologically a
single-contact rf SQUID whose Josephson junction is replaced by a
single-Cooper-pair transistor.\cite{Zorin,Krech2,Friedman,Zorin2}
The latter consists of two mesoscopic junctions with Josephson
energies $E_{J1}, E_{J2}$ and capacitances $C_{1}, C_{2}$, coupled
by a small island. The Hamiltonian of the charge-phase qubit in the
eigenstate basis within the two-level approximation ($E_{CP}\gg
\varepsilon_J(\delta)$) and the small-inductance limit reads as
\cite{R2,Zorin,Krech2}
\begin{equation}\label{H-q}
\displaystyle{ %
  \hat{H}_q=\frac{1}{2}\Delta E\,\hat{\sigma}_z=
  \frac{1}{2}[\varepsilon_J^2(\delta)+D^2(n_g)]^{1/2}\hat{\sigma}_z\,,
}
\end{equation}
where the effective Josephson energy
$\varepsilon_J(\delta)=(E_{J1}^2 + E_{J2}^2 +
2E_{J1}E_{J2}\cos\delta )^{1/2}$ is a function of the total phase
bias across both junctions $\delta=\varphi_1+\varphi_2$ ($\varphi_1$
and $\varphi_2$ are phase differences at the individual junctions);
$D\equiv E_{CP}(1-n_g)$, parameter $n_g=C_gV_g/e$ characterizes the
polarization charge $en_g$ in the island controlled by the gate
voltage $V_g$ via the charge gate capacitance $C_g$, and
$E_{CP}=(2e)^2/2C_\Sigma$ is the two-electron Coulomb energy of the
island expressed through its total capacitance
$C_\Sigma=C_1+C_2+C_g$ regarding the rest of the system. The
inductance $L_q$ of the qubit loop is assumed to conform to the
small-inductance limit (typically $L_qI_c/\Phi_0 \sim 10^{-2}$,
$\Phi_0=h/2e$ is the flux quantum), so that the total flux through
the qubit loop $\Phi\approx\Phi_e$, $\delta\approx
2\pi\Phi_e/\Phi_0$ and the potential (Josephson) energy of the qubit
is periodic on $\Phi$. Thus, the charge-phase qubit design permits
the implementation of an \emph{in-situ} control of Josephson
coupling energy $\varepsilon_J(\delta)$ by means of external
magnetic flux $\Phi_e$ and the effective charging energy $D(n_g)$ by
gate voltage $V_g$. The current operator of an isolated qubit
corresponding to the Hamiltonian (\ref{H-q}) is given by
\begin{equation}\label{I}
\hat{I}= \frac{2e}{\hbar} \frac{\partial\hat{H}_q}{\partial\delta} =
I_q \hat{\sigma}_z, \,\, I_q(n_g,\delta)=
\frac{e}{\hbar}\frac{E_{J1}E_{J2}\sin\delta}{\Delta E(n_g,\delta)},
\end{equation}
so that its basis states are discriminated by the two oppositely
circulating in the qubit loop supercurrents $I_q$ that correspond to
the ground and the excited qubit energy levels.

In general case, at arbitrary ratio of the Josephson and charging
energies $\varepsilon_J(\delta)/E_{CP}$ (while the two-level
approximation (\ref{H-q}) is determined by
$\varepsilon_J(\delta)/E_{CP}\ll 1$) the energy spectrum and
eigenfunctions of the charge-phase qubit are given, respectively, by
the Bloch bands $E_n(q,\delta)$ and the Bloch wave functions
$|q,\delta; n\rangle$ ($n=0,1,2... $ is the band index and $q=en_g$
the quasicharge) that are numerical solutions of Schr\"odinger
equation with Hamiltonian containing all the charge states and
periodic Josephson potential. \cite{Zorin,Zorin2} Then the lowest
two energy levels $n=0, 1$ form the basis of the charge-phase qubit,
so that $ \hat{H}_q=(1/2)\Delta E\,\hat{\sigma}_z,\, \Delta E =
E_1(n_g,\delta)-E_0(n_g,\delta)$.

In our experiments the qubit is driven by coupling microwave field
of frequency $\omega_0$ and amplitude $V_r$ to the voltage gate with
constant voltage component $V_{g0}$, $V_g=V_{g0}+V_r\cos\omega_0 t$,
and its dynamical properties are probed by the well-known rf SQUID
impedance measuring technique\cite{IMT1,IMT2} (IMT) following the
conception of weak continuous quantum
measurements.\cite{Smirnov1,Korotkov,Averin1,Il'ichev,Sillanpaa} In
the IMT the qubit is inductively coupled to a high-quality tank
circuit serving as a linear detector. For the tank circuit driven by
a small amplitude external rf-bias current $I_{rf}\cos\omega_{rf}t$
the measured output functions are the amplitude of the voltage
oscillations $V_T$ and the voltage-current phase shift $\alpha_T$.
Also, the spectrum of voltage fluctuations $S_V(\omega)$ in the
undriven tank can be sensitive to qubit dynamics (the noise
spectroscopy method).

The IMT weak continuous measurements were effectively used for
microwave spectroscopy and characterization of charge-phase
qubits.\cite{CH,CH1} There the resonant response of the qubit to
microwave irradiation was observed through a change of the average
qubit Josephson inductance making contribution to the reactive part
of the tank impedance. Essentially different effects were predicted
in Ref.\,\onlinecite{Smirnov1}, when at resonant microwave
irradiation there exist coherent low-frequency (Rabi) oscillations
in a qubit with frequency nearby the tank resonant frequency. Then
damping and amplification of the voltage signal $V_T$, with
respective modification of the spectrum of voltage fluctuations
$S_V(\omega)$ in the undriven tank and its quality factor, should be
observed owing to insertion of the positive or negative dynamic
active impedance (resistance) into the tank circuit from a qubit.
These phenomena enable probing the coherent properties of qubits
necessary for quantum manipulations.

In the presence of the qubit, at resonant condition
$\omega_{rf}=\omega_T$ of driving the tank, the expressions for its
effective damping rate $\overline{\gamma}_T$, the resonant frequency
$\overline{\omega}_T$, the amplitude of the voltage oscillations
$V_T$ and the voltage-current phase shift $\alpha_T$ have the form
\cite{Smirnov1,Krech1}
\begin{subequations}\label{tank}
\begin{align}
& \displaystyle{ %
\overline{\gamma}_T = \gamma_T + k^2 L_q I_q^2 \omega_T
\chi_{zz}''(\omega_T)\,, } \vspace{0.5cm}\\ %
& \displaystyle{ \overline{\omega}_T =
\omega_T\sqrt{1-k^2L_qI_q^2\chi_{zz}'(\omega_T)}\,, } \vspace{0.5cm}\\
& \displaystyle{ V_T= \frac{I_{rf}}{C_T}\frac{1} {\sqrt{
[k^2L_qI_q^2\omega_T\chi_{zz}'(\omega_T)]^2 +
\overline{\gamma}_T^{\,2}  }}\,,   }\vspace{0.5cm}\\
& \displaystyle{ %
\tan\alpha_T= - k^2 L_q I_q^2 \frac{\omega_T}{\overline{\gamma}_T}
\chi_{zz}'(\omega_T)\,, }
\end{align}
\end{subequations}
where $\gamma_T$ is the own damping rate of the tank,
$k=M/\sqrt{L_TL_q}\ll 1$ is the coupling coefficient, $M$ is the
mutual qubit-tank inductance, $L_T$ and $C_T$ are the inductance and
capacitance of the tank respectively, and $\chi_{zz}'(\omega),
\chi_{zz}''(\omega)$ are the real and imaginary parts of the qubit
magnetic susceptibility $\chi_{zz}(\omega)$. The latter is defined
by the relation
$(d\langle\hat{I}\rangle/dt)_\omega=\chi_{zz}(\omega)MI_q^2
(dI_T/dt)_\omega $ and describes low-frequency response of a qubit
to a tank probing signal. The rate of absorbtion of weak signal
energy by a qubit from a tank $\quad U(\omega)
=(1/2)\omega\chi_{zz}''(\omega)(MI_qI_T)^2 $ ($I_T=V_T/\omega L_T$
is current in the tank), according to properties of generalized
susceptibility.\cite{LLP} That is the function $\chi_{zz}''(\omega)$
defines the absorption ($\chi_{zz}''(\omega)>0$) and irradiation
($\chi_{zz}''(\omega)<0$) properties of a driven qubit coupled to rf
tank-detector.

Renormalization (\ref{tank}a), (\ref{tank}b) of the tank parameters
$\overline{\gamma}_T, \overline{\omega}_T$ at resonance
($\omega=\omega_T$) involves renormalization of the tank quality
factor
\begin{equation}\label{Q}
\begin{array}{l}
\displaystyle{ %
Q=\frac{\overline{\omega}_T }{\overline{\gamma}_T}=
\frac{\overline{\omega}_T \overline{L}_T}{\overline{R}_T}=
\frac{\overline{\omega}_T (L_T+L_q)}{R_T+R_q},\,\,
\overline{\omega}_T=(\overline{L}_T C_T)^{-1/2}\,, } \vspace{0.3cm}\\ %
\displaystyle{ %
L_q\approx L_T\,k^2L_qI_q^2\chi_{zz}'(\omega_T),\,\, %
R_q\approx \omega_T L_T\,k^2L_qI_q^2\chi_{zz}''(\omega_T), }
\end{array}
\end{equation}
where $\overline{L}_T = L_T+L_q$ is the effective inductance of the
tank including inserted from the qubit inductance $L_q$, and
$\overline{R}_T = R_T+R_q $ is the effective resistance of the tank
including inserted from the qubit resistance $R_q$ [$R_q(\omega)=\text{Re}Z(\omega)$, %
where $Z(\omega)$ is qubit impedance seen by the tank]. %
The characteristic impedance of the tank at resonance
$$ R_c=\frac{V_T}{I_{rf}}= \frac{Q}{\overline{\omega}_T C_T}=
Q^2 \overline{R}_T = \frac{\overline{L}_T}{\overline{R}_T C_T}\,, $$
so that the tank voltage $V_T$ is inversely proportional to the
effective tank resistance $\overline{R}_T$. The Lorentzian-shaped
voltage fluctuations spectrum $S_V(\omega)$ in the tank is
characterized by the renormalized quality factor (\ref{Q}).

For the special case of charge-phase qubits, the theory of Rabi
oscillations and low-frequency qubit response have been investigated
in Ref.\,\onlinecite{Krech1}. The Rabi frequency $\Omega_R$ of
coherent low-frequency oscillations generated at near-resonant
driving of the charge-phase qubit by microwave field reads
as\cite{Krech1}
\begin{equation}\label{OR}
\begin{array}{l}
\displaystyle{ %
  \Omega_R=\sqrt{\Omega_0^2+\varepsilon^2},\,\,\,
  \varepsilon=\omega_g-n\omega_0,\,\omega_g=
  \Delta E/\hbar,\,|\,\varepsilon\,|\ll \omega_0,} \vspace{0cm}\\
\displaystyle{ %
  \Omega_0=n\omega_0\tan\eta\,J_n(\alpha),\,\,
  \alpha=\frac{2eV_r}{\hbar\omega_0}\frac{C_g}{C_\Sigma}\cos\eta },
\end{array}
\end{equation}
where $\Omega_0$ is the core $n$-photon Rabi frequency,
$\tan\eta=\varepsilon_J(\delta)/D(n_g)\lesssim 1$. The most
essential in our context low-frequency parts of the functions
$\chi_{zz}'(\omega), \chi_{zz}''(\omega)$ in the presence of Rabi
oscillations have the form\cite{Krech1}
\begin{equation}\label{hi}
\begin{array}{l}
\displaystyle{ %
\chi_{zz}'(\omega)= \frac{P}{\hbar}
\left(\frac{\Omega_0}{\Omega_R}\right)^2
\frac{\Omega_R-\omega}{ (\omega-\Omega_R)^2 + \Gamma^2 }\,, } \vspace{0.1cm} \\
\displaystyle{ %
\chi_{zz}''(\omega)= \frac{P}{\hbar}
\left(\frac{\Omega_0}{\Omega_R}\right)^2 \frac{\Gamma}{
(\omega-\Omega_R)^2 + \Gamma^2 }\,, }
\end{array}
\end{equation}
where $\Gamma$ denotes the dephasing rate for the transverse
component of the Bloch vector to determine the qubit dynamics, the
polarization constant $P= (\varepsilon / \Omega_R) \langle
\hat{\sigma}_z \rangle$, $\langle \hat{\sigma}_z
\rangle=\mbox{Sp}\{\hat{\rho}\hat{\sigma}_z\}=1-2P_+,$ $P_+ \leq
0.5$ being the probability of the qubit excited state ($\hat{\rho}$
the reduced density matrix of the qubit), at that $0\le\langle
\hat{\sigma}_z \rangle\le 1$. It is essential that the parameter
$P$, and therefore $\chi_{zz}''(\omega)$ and the qubit-induced
contribution to the effective damping rate of tank
$\overline{\gamma}_T$, (\ref{tank}a), alter their sign when the
detuning parameter $\varepsilon$ does so. Thus the effective
resistance $\overline{R}_T \propto \overline{\gamma}_T$ contains a
negative contribution $R_q$ from the qubit at $\omega_g(\Phi_e)<
n\omega_0$ ($\varepsilon <0$, $\overline{\gamma}_T<\gamma_T$) and a
positive contribution at $\omega_g(\Phi_e)> n\omega_0$ ($\varepsilon
>0$, $\overline{\gamma}_T>\gamma_T$).

As seen from (\ref{hi}), the functions $\chi_{zz}'(\omega),
\chi_{zz}''(\omega)$ have non-monotonous resonant character near
$n$-photon Rabi frequencies. In their turn, as follows from
(\ref{tank}), the weak continuous quantum measurement output
functions $V_T, \alpha_T$ of the tank-detector expressed through
$\chi_{zz}'(\omega), \chi_{zz}''(\omega)$ have resonant
contributions, conditioned by coherent Rabi dynamics of the qubit at
$\Omega_R\simeq \omega_T$ with small damping and decoherence rates,
$\Gamma < \Omega_R$. Notice that in previous spectroscopy
experiments with charge-phase qubits\cite{CH,CH1} there was only
small changing of the tank effective resistance ($\Gamma > \Omega_R$
and $\chi_{zz}''(\omega)\approx 0,\, \overline{R}_T\approx R_T$) and
the voltage $V_T$ conditioned by $\chi_{zz}'(\omega)$ could only
decrease, as seen from (\ref{tank}c). In case of coherent Rabi
dynamics at $\Omega_R\approx \omega_T$, $\chi_{zz}''(\omega)$ can be
nonzero while $\chi_{zz}'(\omega)\approx 0$, and the voltage $V_T$
can both decrease or increase relative to the voltage level of the
passive tank depending on sign of $\chi_{zz}''(\omega_T)\propto
R_q$.

Recently, quantum theory of the low-frequency linear susceptibility
of flux and charge-phase qubits 
subjected to microwave irradiation was developed within the density
matrix formalism in Ref. \onlinecite{Greenberg}. This theory as well
describes parametric effects of reduction and amplification of
signal and quality of the tank coupled to charge-phase qubits in the
strong charge limit $E_{CP}\gg \varepsilon_J(\pi)$.

The above theoretical outline gives only a qualitative picture of
the effects in the system of charge-phase qubit and tank-detector
and cannot afford for their quantitative description. %
Firstly, the charge-phase qubit has such parameters that
$E_{CP}\gtrsim \varepsilon_J(\pi)$ in our experiment, while its
Hamiltonian and the energy spectrum are of more complicated
structure as compared to the two-level approximation (\ref{H-q}),
and numerical solving the Schr\"odinger equation is needed for this
structure to be found. Another important feature is a strong
nonlinearity of the qubit's response to external electromagnetic
excitation due to Josephson character of the energy potential for
the quantum coordinate of the system. As shown in the
quantum-mechanical model of a SQUID ring coupled to an
electromagnetic field,\cite{Everitt} the energy levels of the SQUID
with coherent dynamics of magnetic flux when excited by strong
electromagnetic field change so dramatically in time that their time
averages essentially differ from the initial system levels. The
shape of the time-averaged energy levels $\langle
E_0\rangle(\Phi_e), \langle E_1\rangle(\Phi_e)$ in the high-power
electromagnetic field depends on its frequency and amplitude in a
complicated manner showing compression between the levels and the
interference patterns of resonant interactions between the two
lowest ones. Similar by their nature effects are observed in
nonlinear quantum optics. So, a two-level atom in strong
electromagnetic field demonstrates nonlinear effects of compression
its energy (quasi)levels and Stark effect.\cite{Krainov} Another
basic aspect of the theoretical consideration of quantum dynamics of
the qubit-detector system is taking into account the temperature
effect resulting in finite width of the signal generation band in
the qubit and in the tank, thus determining eventually the shape of
the observed signal characteristics.

\section{Charge-phase qubit and experimental results}
\subsection{Qubit calibration}

The sample (see Fig.\,\ref{fig01}) was fabricated by electron beam
lithography and conventional two-angle technology of aluminum thin
film deposition to create two mesoscopic Josephson junctions with
area $S_{J1,J2}\approx 3.5\!\cdot\! 10^4\,$nm$^2$, tunnel resistance
$R_{J1,2}\approx 3.0\,\mbox{k}\Omega$ and $E_{J1,2}/h\simeq
53\,$GHz. The evaluation of $R_{J1,2}$ was obtained from direct
measurements of the tunnel resistance of a large-area "witness".
Analysis of the scanning electron micrograph of the sample shows
that the areas of the Josephson junctions in this charge-phase
sample differ by appr. $15\,\%$. The difference between the contacts
areas determines the asymmetry of the critical currents and leads to
the optical evaluation of minimal value of the effective Josephson
energy $\varepsilon_J(\pi)/h=|E_{J1}-E_{J2}|/h\approx 7\,\mbox{GHz}$
of the Cooper-pair transistor. The two-electron Coulomb energy
evaluated on the basis of the geometric transistor capacitance is
$E_{CP}/h \approx 17\,\mbox{GHz}$. Real value of $E_{CP}$ may be
somewhat smaller due to effect of the stray capacitance of the
island. Thus, the two characteristic energies in the given qubit are
of the same order of magnitude and meet the adiabaticity condition
$\varepsilon_J(\pi)< E_{CP} < \Delta_{Al}(0)$.

Choosing $E_{CP}\gtrsim \varepsilon_J(\pi)$ instead of strong
condition $E_{CP}\gg \varepsilon_J(\pi)$ we decrease the influence
of nonequilibrium noise of the charge on coherent dynamics of the
qubit.\cite{R2} With the same aid,\cite{Zorin2,Makhlin} both the
ratio $(C_g/C_\Sigma)^2 \approx2\!\cdot\!10^{-3}$ and the ratio of
the real part of the charge-gate line impedance to the quantum
resistance $\mbox{Re}Z_g(\omega)/R_Q\approx 10^{-2}$, where
$R_Q=h/4e^2$, are made small in the given sample of charge-phase
qubit. To reduce the qubit effective temperature $T_{eff}$, all the
input electric circuits were thoroughly filtered by multiresonance
(Cu)O powder filters, and the voltage supply resistor along with the
last stage of the copper filter of the charge-gate line were placed
onto $10$\,mK refrigerator stage for better thermalization.

The island volume permits  one making an estimation of the number of
states $N_{eff}$ and the characteristic temperature\cite{Tuominen}
$T^{\ast}=\Delta_{Al}(0)/k_B \ln N_{eff}\simeq 120\,$mK above which
the number of the quasiparticle excitations in the island abruptly
rises from zero.

The charge-phase qubit Hamiltonian (\ref{H-q}) is sensitive to any
changes in magnetic flux coupled to the superconducting loop. To
minimize the external flux noise, the qubit superconducting loop is
implemented as a 1st order gradiometer with minimum distance
between the coils. The achieved balancing accuracy (difference
between the areas of the gradiometer coils) is of the order of one
part in $10^{3}$.

To read the change in the amplitude $V_T$ and phase $\alpha_T$ of
the oscillations, the tank circuit was linked to a preamplifier
placed on $1\,$K-pot and room temperature electronics. It should be
emphasized that the part of the tank circuit kept at temperature
$T\approx 1.7\,$K ($1\,$K-pot) (see Fig.\,\ref{fig01}) is a source
of high-frequency photons that can affect the qubit. Moreover, in
addition to amplifying the in-band Rabi-signal, it is obvious that
preamplifier can also irradiate at extremely high frequencies thus
increasing the effective temperature $T_{eff}$ of the qubit. The
qubit decoherence due to back-action from both the tank circuit and
the preamplifier depends on many parameters and was not analyzed yet
in detail. We will return to this problem below. To reduce
$T_{eff}$, the coupling coefficient was decreased to $k\approx
0.0295$ that led to the parameter $k^2Q\approx 0.3$. However,
coupling to the tank circuit cannot be made arbitrarily small in
Rabi qubit-based detectors for the purpose of reducing decoherence
and should be optimized. We have found that $k^2Q\simeq 0.3$ is
close to the most optimal value that provides maximum
(Rabi)signal-to-noise ratio due to effect of the high-frequency
photons from the measuring circuit. This value substantially differs
from the usual match condition $k^2Q\gtrsim 1$ for classic rf
SQUIDs.\cite{Likharev,IMT1,IMT2} Despite the main part of the tank
circuit is placed at $T\simeq 10\,$mK, its effective noise
temperature $T_T^N\simeq 400\,$mK in our experiments. It can be
shown\cite{Smirnov1,Zorin2} that at $\Omega_R=\omega_T$ the tank
with impedance $\omega_T L_T Q\simeq11\,$k$\Omega$ at mutual
inductance $M=k\sqrt{L_TL_q}\simeq0.365\,$nH in the band
$\gamma_T=85\,$kHz contributes sufficiently small noise flux
$\Phi_T^N \sim 2\!\cdot\! 10^{-6}\Phi_0$ to the qubit outside the
region of exact resonance.

The previous measurements displayed that superfluous noises from the
microwave line at galvanic coupling with a qubit were unavoidable.
Therefore all experimental results were obtained with the qubit
placed in a Pb resonator in the region of maximum electric field.
The qubit was then continuously irradiated with a microwave field to
introduce an alternating voltage component $V_r\cos\omega_0 t$. So,
the capacitance $C_E$ formally galvanicly coupled to the charge gate
capacitance $C_g$ in Fig.\,\ref{fig01} denotes, in fact, coupling of
the qubit charge gate to a microwave field. Since the distribution
of the electric field only approximately corresponds to that
expected for an ideal resonator, we made preliminary measurements of
the qubit response to the microwave field as a function of its
frequency in the range of 6.6 to 8.0\,GHz.

The detailed analysis of the qubit dynamics has shown that extremal
positive (peak) and negative (dip) responses $V_T(\Phi_e)$ to
low-amplitude microwave field was observed at frequency
$f\equiv\omega_0/2\pi\cong 7.27\,$GHz, with small deviations
$0.010\Phi_0$ and $0.017\Phi_0$ from the point $\Phi_e=0.5\Phi_0$.
Assuming that the single-photon process of the higher-level
excitation occurs at frequency $f=7.27$\,GHz, we find the effective
Josephson energy $\varepsilon_J(\pi)/h\approx 6.4\,$GHz.

\subsection{Single-photon Rabi response}
\begin{figure}[t!]
\centering %
\scalebox{0.99}[0.99]{\includegraphics[bb = 15 15 271 335]{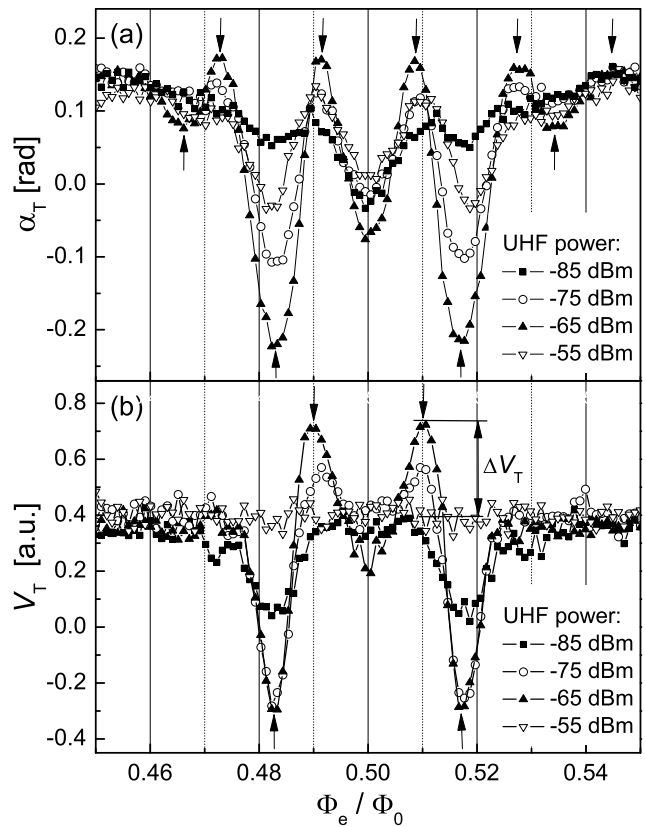}} %
\vspace{-0.5cm} %
\caption{\label{fig02} Experimental results for the
charge-phase qubit placed in the region of the maximum electric
field under continuous microwave irradiation with $f=7.27\,$GHz. a)
Set of the curves of the voltage-current phase shift
$\alpha_T(\Phi_e/\Phi_0)$ in the tank circuit, b) Set of the curves
of the amplitude of the voltage oscillations $V_T(\Phi_e/\Phi_0)$ in
the tank circuit. The curves parameter is the microwave power. The
arrows denote extrema of $\alpha_T(\Phi_e/\Phi_0),
V_T(\Phi_e/\Phi_0)$ for single-, two- and three-photon excitations.
The signal shape near resonances is determined by the finite width
of the Rabi generation band. The maximum response is observed at
$P\approx -65\,$dBm. }
\end{figure}

Assumed that one should observe resonant parametric
response of the qubit in the region of Rabi frequencies $\Omega_R
\simeq \omega_T$ when the microwave field frequency was near that of
single-photon transitions, we investigated in detail dependencies
$V_T(\Phi_e)$ and $\alpha_T(\Phi_e)$ at various amplitudes (powers)
of the microwave field. Typical measurement results at $f=7.27\,$GHz are
shown in Figs.\,\ref{fig02}a,b. It is clearly seen from these curves
that the single-photon resonant response predicted by
(\ref{tank})\,--\,(\ref{hi}) and smeared out by temperature noise
and coupling to the readout device is observed at an optimal
microwave power. The decrease, as well as increase, of the microwave
power compared to its optimal value results in falling of the
resonant response amplitude that is explained by the rise of the
detuning $\xi=\Omega_R(V_r)-\omega_T$.

The variations $V_T(\Phi_e), \alpha_T(\Phi_e)$ in these experiments
are measured simultaneously and, since the rf generator frequency is
chosen from the condition $\omega_{rf}\cong\omega_T$, then the
technique sensitivity is maximal to the change in the oscillations
phase. On the contrary, sensitivity of the voltage channel to small
changes of the resonant frequency of the tank $\omega_T$ near
$\omega_T\approx\omega_{rf}$ is minimal. Notice the two-photon
resonant response seen fairly well in $\alpha_T(\Phi_e)$ curves in
the vicinity of $\Phi_e=0.53\Phi_0$ (Fig.\,\ref{fig02}a) whereas
having signal-to-noise ratio $S/N\approx 1$ at $V_T(\Phi_e)$
characteristics. The decrease of the amplitude of the two-photon
resonant response relative to the single-photon one in
Fig.\,\ref{fig02}a agrees with the theory\cite{Smirnov1} that
predicts the longest decoherence time in the point of symmetry
$\Phi_e=0.5\Phi_0$.

In general case, variations of the phase as well as the amplitude of
the parametric oscillations in the tank circuit are bound to the
change of its resonance frequency and quality. Thus, presented in
Fig.\,\ref{fig02}a,b dependencies $\alpha_T(\Phi_e)$, $V_T(\Phi_e)$
qualitatively confirm existence of the low-frequency part of the
charge-phase qubit susceptibility functions (\ref{hi}) as the Rabi
frequency $\Omega_R$ of the qubit passes through the resonant
frequency $\omega_T$ of the tank. Let us consider the evolution of
the signals $V_T$ in vicinity of the point $\delta=\pi$. As it
follows from the expressions (\ref{tank}), Rabi-response of the
qubit in the point $\delta=\pi$ should vanish to zero proportionally
to the square of the circulating current $I_q^2$. At low amplitudes
of the microwave field the responses $V_T$ are really small as it is
seen from Fig.\,\ref{fig02}b. However, at power $P=-65\,$dBm
noticeable shifts $\Delta V_T$ (as well as $\Delta\alpha_T$) are
observed that points out to appearance of circulating current in the
point $\delta=\pi$ in the expense of deviation of time-averaged
eigenenergies from stationary eigenenergies.\cite{Everitt}

\begin{figure}[t!]
\centering %
\scalebox{0.92}[0.92]{\includegraphics[bb = 15 15 306 211]{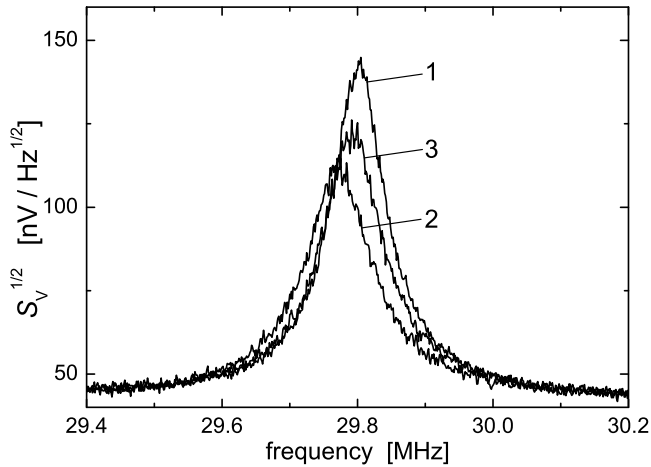}} %
\vspace{-0.5cm}%
\caption{\label{fig03} Noise spectroscopy. Set of curves of noise
spectral density of the tank circuit as a function of frequency
obtained at optimal microwave power ($P\approx -65\,$dBm) in extreme
points: {\it 1} --- $\Phi_e^{peak}\approx \Phi_0(0.5\pm 0.010)$
($Q^{peak}\approx 360$), {\it 2} --- $\Phi_e^{dip}\approx
\Phi_0(0.5\pm 0.017)$ ($Q^{dip}\approx 260$), and in the degeneracy
point {\it 3} --- $\Phi_e\approx 0.5\Phi_0$ ($Q\approx 301$). The
resonant frequencies $\omega_T/2\pi$ in these points are: 29.803,
29.770, 29.795\,MHz respectively. }
\end{figure}

We use the noise spectroscopy method for the direct impedance
measurements. In this case the generator current $I_{rf}=0$, and the
tank circuit is excited by thermodynamic fluctuations while its
characteristics depend on the impedance inserted from the qubit
under continuous irradiation. Fig.\,\ref{fig03} displays the noise
spectral density as a function of frequency for the two extreme
("peak"\, and "dip") points denoted by arrows in Fig.\,\ref{fig02}b
and at the degeneracy point $\delta=\pi$. It is seen from these
amplitude-frequency characteristics measured at minimal influence of
the tank circuit on the qubit that Rabi oscillations result in
visible change of the tank quality and shift of the tank resonant
frequency in the "peak"\, and "dip"\, $\Phi_e$-points (curves {\it
1} and {\it 2} in Fig.\,\ref{fig03}, respectively). The sign of the
inserted from the qubit Rabi-impedance is determined by the sign of
the detuning $\varepsilon=\omega_g-\omega_0$ between energy gap
$\omega_g(\Phi_e)$ and microwave field frequency $\omega_0$. From
the quality factor values and the resonant frequencies in the
extreme points indicated in Fig.\,\ref{fig03} and using (\ref{Q}),
we will obtain that the maximum value of the negative resistance
inserted into the tank $R_q=R^{peak}\approx -0.0173\,\Omega$
practically coincides with the maximum positive resistance
$R^{dip}\approx 0.0176\,\Omega$ by absolute magnitude while the
alteration of its sign implies the transition from the irradiation
to the absorption properties of the qubit. The inductance
contributions in these points are small ($L_q^{peak}\approx
-0.53\!\cdot\!10^{-3}L_T, L_q^{dip}\approx
1.68\!\cdot\!10^{-3}L_T$) and are also of opposite signs. %
Note that similar electrodynamics is typical for the parametric
interaction of a resonator microwave signal with Josephson
oscillations.\cite{Kanter} In this classic analogue, the maximum
amplitude of the parametric effect is determined by fluctuations of
the average contact voltage that lead to broadening of the Josephson
generation band. So, assuming that the rise of the qubit temperature
in the considered case of Rabi oscillations will result in
broadening of their spectrum, one should expect some decrease in
amplitude of the inserted negative impedance with increasing the
bath temperature.

\begin{figure}[t!]
\centering %
\scalebox{0.92}[0.92]{\includegraphics[bb = 15 15 301 217]{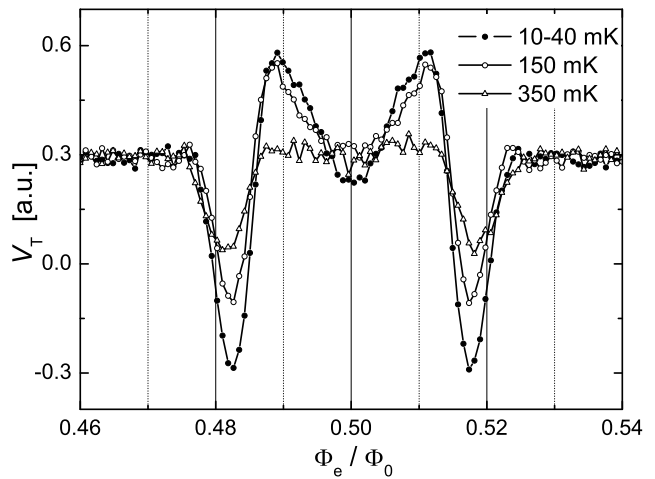}} %
\vspace{-0.7cm}%
\caption{\label{fig04} Set of curves $V_T(\Phi_e/\Phi_0)$ obtained
under microwave irradiation with frequency $\omega_0=7.27\,$GHz.
The microwave field amplitude is optimal for each curve. The curve
parameter is the refrigerator temperature. }
\end{figure}

The analysis of temperature dependencies demonstrates that the
amplitude of the inserted resistances $R^{peak}, R^{dip}$ within
experimental accuracy does not depend on $T_{bath}$ in the interval
$T_{bath}=\,$10 -- 40\,mK. The amplitude of the inserted impedance
decreased by $15\,\%$ at $T_{bath}=60\,$mK and the extrema in curves
$V_T(\Phi_e/\Phi_0)$ smooth out in the interval 60 -- 100\,mK . The
increase of the bath temperature up to 150 -- 200\,mK leads to the
dramatic fall of the amplitude $\Delta V_T$. Note that the
calculated temperature $T^{\ast}\cong 120\,$mK, at which abrupt
decay of the coherent effects is expected, is close to the
experimental data. The coherence degradation at $T>T^{\ast}$ is
accompanied by fast shortening of the decoherence time and widening
of the Rabi oscillation band. Further temperature $T_{bath}$ rise up
to 300 -- 350\,mK causes Rabi-impedance of the qubit seen by the
tank vanishing to zero ($\Delta V_T=0$) within the noise level, i.e.
the coherent Rabi effects disappear.

The minimal effective temperature of the qubit estimated to be
$T_{eff}\approx 80\,$mK is determined mainly by the effect of the
readout circuit (tank-detector and HEMT amplifier) and leads to
sufficiently large uncertainty in the magnetic flux
$\delta\Phi\simeq(k_BL_qT_{eff})^{1/2}\simeq 0.01\Phi_0$ for the
qubit under consideration. Anyway, the dependence
Re$Z(\omega)\propto \chi_{zz}''(\omega)$ on the microwave power and
the temperature indicates the presence of coherent quantum
oscillations at low temperatures that have potential applications as
a self-Rabi pumping signal in quantum information detectors.

It is seen from the results presented in Fig.\,\ref{fig04} that at
the bath temperature 350\,mK near $\Phi_e^{dip}\approx\Phi_0(0.5\pm
0.017)$ the "dips" are yet observed and associated with the
processes of noncoherent quantum tunneling of Cooper pairs, without
forming a superposition of the charge states. Therefore, the
contribution of the coherent oscillations in the given temperature
interval is described as the difference between the values of
$V_T(\Phi_e)$ obtained at 10 and 350\,mK. Taking into account that,
at optimal amplitude of the microwave field, presence of the
coherent oscillations leads to $20\%$ change in the tank quality,
the evaluation can be made at our parameters for the minimum
decoherence time $\tau_\varphi^{min}>2\pi/\omega_T \approx
0.033\,\mu$s. While the charge-flux qubit is an ideal frequency
downshifter, one can evaluate then $\tau_\varphi$ from the
high-frequency resonance linewidth that yields 3 times as large
values of $\tau_\varphi \simeq 0.1\,\mu $s.

According to the theory,\cite{Krainov} the enhancement of $Q_q$ is
possible not only at the expense of lowering the noise temperatures
of the qubit and the tank but also due to narrowing the $n$-photon
resonance band when going to the region of the multi-photon Rabi
oscillations. The evidence of the two- and three-photon resonances
in a similar charge-phase qubit were demonstrated in previous
work.\cite{CH1}

\subsection{Multi-photon Rabi response}

It follows\cite{Krainov} from the theory of two-lewel atom in strong
low-frequency electromagnetic field that the quality of multi-photon
resonance considerably rises while the resonance order grows owing
to coherent addition of individual transition amplitudes. The
narrowing of distribution of the excited level occupation
probability must result in a natural filtration of noise photons in
the discussed-above effect of the interaction between the tank
circuit current and the Rabi-like oscillations in the charge-phase
qubit. At the same time, the amplitude of the synchronization effect
causing the energy irradiation (absorption) by the qubit at
frequencies close to that of the multi-photon Rabi oscillations may
increase. In other words, the signal characteristics, the conversion
steepness and the sensibility of a parametric detector based on
charge-phase qubit must enhance when coming to the multi-photon
excitation. Depending on the interplay between quantities included
in Hamiltonian (\ref{H-q}), such a detector could be sensitive to
the induced charge like rf-SET detector,\cite{SET-detector} and to
the magnetic flux like SQUIDs\cite{IMT1,IMT2} and detectors based on
the flux qubit.\cite{Flux-detector}

\begin{figure}[t!]
\centering %
\scalebox{1.}[1.]{\includegraphics[bb = 15 15 267 230]{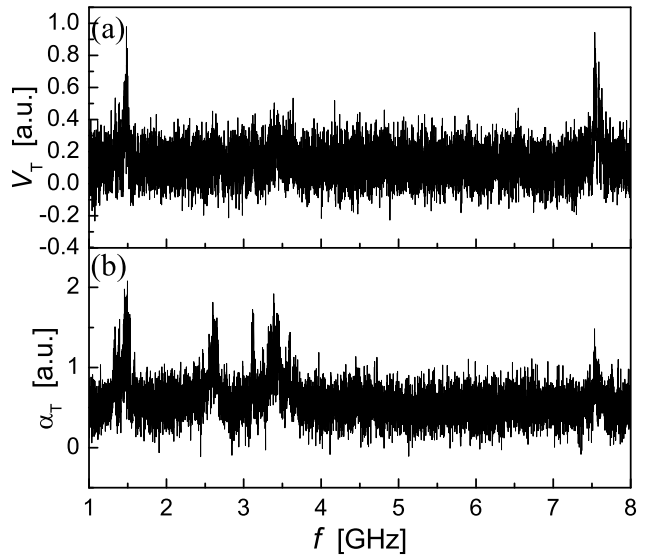}} %
\vspace{-0.5cm}%
\caption{\label{fig05} Resonant excitation of the charge-phase
qubit. Amplitude $V_T(f)$ and phase $\alpha_T(f)$ of the tank
circuit oscillations ($M I_T<10^{-4}\Phi_0,
\omega_{rf}\approx\omega_T$) vs. frequency of the microwave field in
the charge gate of the charge-phase qubit with $n_g\approx 1,
\delta\approx\pi$. The 1/f noise level becomes important because of
long-continued (4 hours) spectrum registration. Multi-photon
resonance from the ground state to the first excited state induced
by the photon frequency $f\approx 1.4\,$GHz is clearly seen. The
generator microwave field amplitude is kept constant $P=-60\,$dBm
when frequency sweeping, the bath temperature being $10\,$mK. %
}
\end{figure}

Spectroscopic studies were performed in wide (1 -- 8\, GHz)
frequency range to find the frequencies of the multi-photon
resonances of the charge-phase qubit ($n_g\approx 1,
\delta\approx\pi $). At sweeping the frequency of small-amplitude
microwave field the time-averaged upper level occupation probability
goes through a pronounced maximum at frequency values meeting to the
exact multi-photon resonance condition\cite{MFLZ} that lead to
maximum qubit response. However, the effectiveness of excitation of
the multi-photon resonance depends on the generator power and the
distribution of the field in the resonator. So, if for a certain
frequency of the multi-photon resonance the charge-phase qubit is
situated in a peak of the electrical field then the amplitude
$V_r\cos\omega_0 t$ at the charge gate reaches its maximum.

During the multi-photon experiments, the parameters of the
charge-phase qubit, its position inside the resonator and the
measuring circuitry (Fig.\,\ref{fig01}) to detect the response by
the technique of weak continuous measurement remained the same, as
in "single-photon" measurements described in previous section.

\begin{figure}[t!]
\centering %
\scalebox{0.95}[0.95]{\includegraphics[bb = 15 15 280 563]{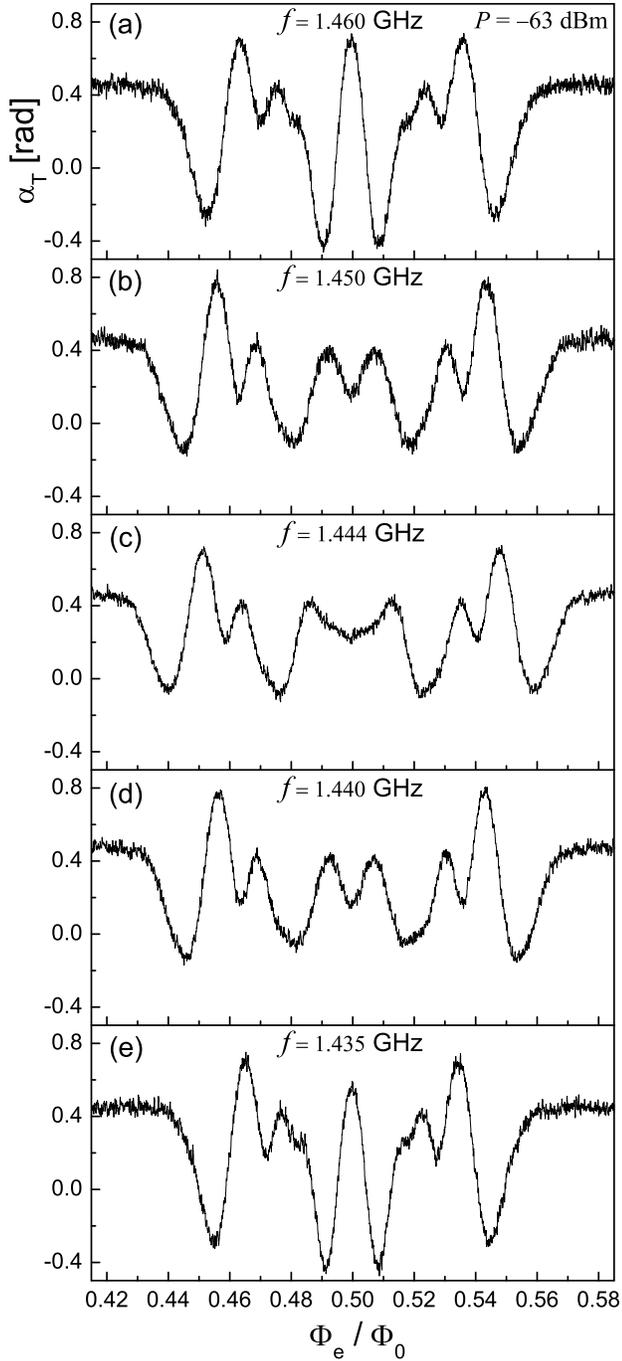}}%
\vspace{-0.2cm}%
\caption{\label{fig06} Set of signal characteristics
$\alpha_T(\Phi_e/\Phi_0)$ of charge-phase qubit in microwave field
in the vicinity of five-photon resonance at the bath temperature
10\,mK. The curves parameter is the microwave frequency. The
generator power $P=-63\,$dBm is kept constant for all the curves.
The characteristics set (a-e) reflects the dependence of the
probability excitation on the frequency at positive (a), (b) and
negative (d), (e) detuning relative to resonance frequency
$f=1.444\,$GHz (c). %
}
\end{figure}
\begin{figure}[t!]
\centering %
\scalebox{0.95}[0.95]{\includegraphics[bb = 15 15 280 360]{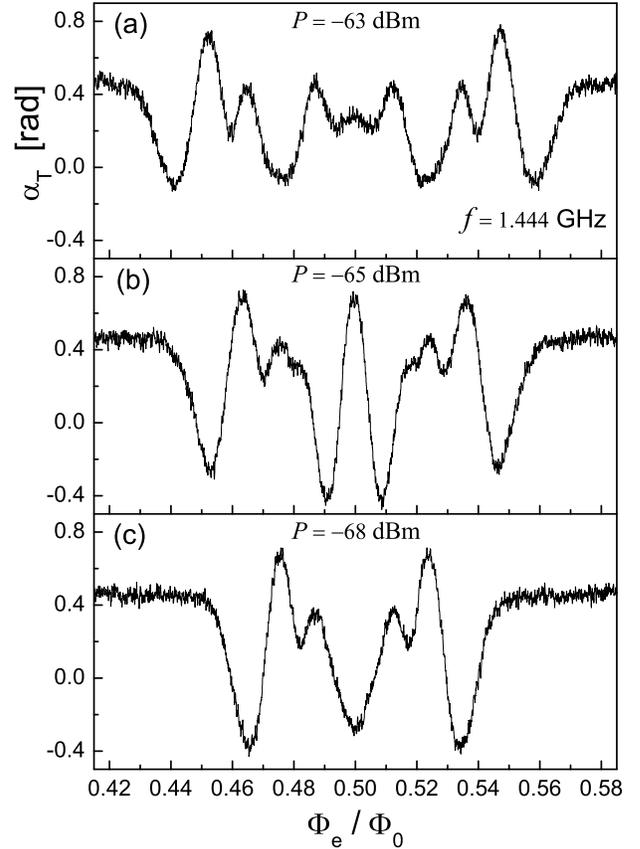}} 
\vspace{-0.6cm}%
\caption{\label{fig07} Set of signal characteristics
$\alpha_T(\Phi_e/\Phi_0)$ of charge-phase qubit in microwave field
with frequency $f=1.444\,$GHz. The curves parameter is the microwave
power. It is well seen that the dependence taken at $P=-65\,$dBm
practically coincides with the curves obtained at $P=-63\,$dBm and
at frequencies 1.460\,GHz and 1.435\,GHz, see Fig.\,\ref{fig06}a,e.
}
\end{figure}

It follows from the results of the level spectroscopy shown in
Fig.\,\ref{fig05} that under given condition the five-photon
resonance is easily excited at a frequency close to 1.4\,GHz
($7.27\mbox{\,GHz}/1.4\mbox{\,GHz}\approx 5$). Note that the
microwave generator-to-resonator and resonator-to-qubit transmission
lines have their own resonances (frequency-dependent coupling
between the MW-generator and qubit). This evokes a non-monotonic
frequency dependence of the injected microwave power and, with
taking into account the nonlinear transformation in the qubit, to
the appearance of additional spectral lines. Noticeable spread in
the peaks positions along the frequency axis and excessive peak
width in Fig.\,\ref{fig05} are associated with a drift of the qubit
parameters (1/f noise) because of long-lasting ($\sim 4$ hours)
frequency sweeping. Therefore, at the next stage, analyzing the
qubit response $\alpha_T(f)$ to microwave field in a narrow
frequency band, we have found the resonant frequency for five-photon
transition was close to $f=1.444$\,GHz.

\begin{figure}[t!]
\centering %
\scalebox{0.95}[0.95]{\includegraphics[bb = 15 15 285 260]{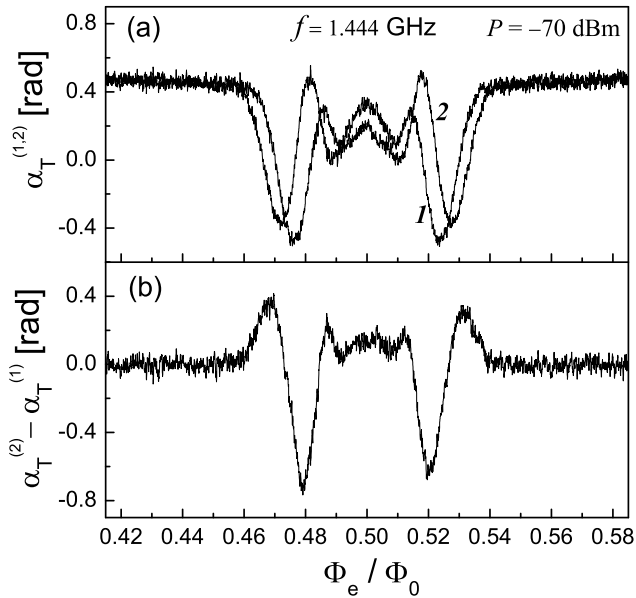}} %
\vspace{-0.5cm}%
\caption{\label{fig08} (a) Signal characteristics
$\alpha_T^{(1,2)}(\Phi_e/\Phi_0)$ of charge-phase qubit in microwave
field ($P=-70\,$dBm, $f=1.444\,$GHz) for two values: $n_g\approx 1$
(1) and $n_g\approx 0$ (2) in the charge gate. (b) The signal
difference $\alpha_T^{(2)}-\alpha_T^{(1)}$ for the change %
$\Delta n_g=n_g^{(1)}-n_g^{(2)}\approx 1$. %
}
\end{figure} %

To determine the effective width of the resonance line (quality
factor), two series of measurements were performed of the phase of
oscillations in the tank circuit as a function of the magnetic flux
applied to the qubit and the microwave field parameters: (i) --
$\alpha_T(\delta)$ at a fixed microwave field amplitude $V_r$ for
several frequencies near $f=1.444$\,GHz and (ii) --
$\alpha_T(\delta)$ at the resonant frequency $f=1.444$\,GHz for
several microwave field amplitudes. Typical results for the
measurements of the series (i) and (ii) are shown in
Fig.\,\ref{fig06} and Fig.\,\ref{fig07}, respectively. Since maximum
detuning of the microwave generator from the resonance for the set
presented in Fig.\,\ref{fig06} is small
$\Delta\omega_0/\omega_0\cong0.014$, the power in the channel can be
considered then as being practically constant while sharp dependence
of the $\alpha_T(\delta)$ shape on the microwave field frequency
should be associated with the quality of the five-photon resonance.
It follows from the comparison between the series of the
measurements (i) and (ii) that drop in the microwave field power by
2\,dB at fixed frequency $f=1.444$\,GHz (couple of curves in
Fig.\,\ref{fig07}a and Fig.\,\ref{fig07}b) results practically in
the same transformation of the $\alpha_T(\delta)$ shape as when the
microwave field frequency changes from $f=1.444$\,GHz  to 1.460\,GHz
or to 1.435\,GHz at a fixed generator power (couple of curves in
Fig.\,\ref{fig06}c and Fig.\,\ref{fig06}a or Fig.\,\ref{fig06}c and
Fig.\,\ref{fig06}e, respectively). Therefore, the effective quality
of the five-photon resonance in the charge-phase qubit reaches
$Q_{n=5}=1.444/(1.460-1.435)\approx 58$ at 2\,dB change of the
microwave power. Taking the two shape-alike dependences at positive
(Fig.\,\ref{fig06}b) and negative (Fig.\,\ref{fig06}d) detuning and
assuming that the population at a multi-photon resonance is almost
symmetric with respect to the detuning sign, we will have for the
resonance frequency $f=(1.440+1.450)/2\approx 1.445$\,GHz that is
close to the value determined earlier.

\begin{figure}[t!]
\centering %
\scalebox{0.95}[0.95]{\includegraphics[bb = 15 15 280 260]{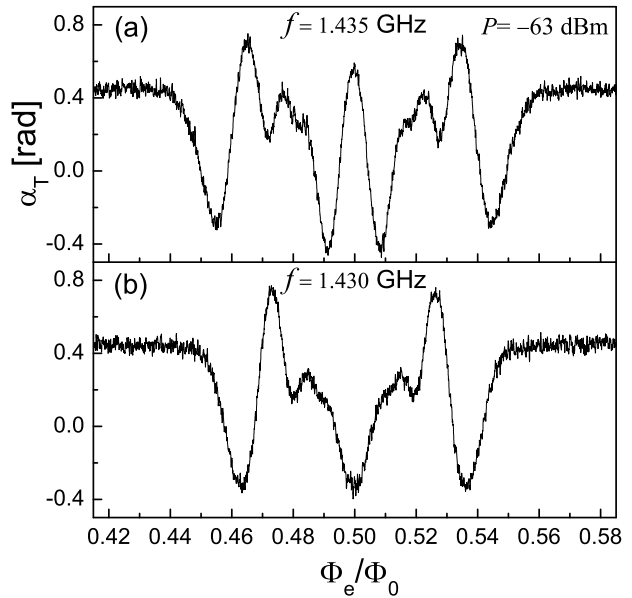}} %
\vspace{-0.5cm}%
\caption{\label{fig09} Interference-induced variation in signal
characteristics $\alpha_T(\Phi_e/\Phi_0)$ of the charge-phase qubit
at the symmetry point $\Phi_e=0.5\Phi_0$. Maximum
$\alpha_T(0.5\Phi_0)\approx 0.5$ (a) transforms into minimum
$\alpha_T(0.5\Phi_0)\approx -0.4$ (b) when the microwave frequency
is detuned by $\Delta f= 5\,$MHz ($\Delta f/f = 0.0035$) at fixed
microwave generator power $P=-63\,$dBm. }
\end{figure}

Owing to the high effective quality of the five-photon resonance,
the response of charge-phase qubit to the charge variation
considerably increases as compared to the case of single-photon
excitation. As it is seen from Fig.\,\ref{fig08}, the change of the
polarization charge by $\Delta n_g\approx 1$ causes
$\Delta\delta\simeq 0.01\pi$ shift of the peaks and dips associated
with the qubit's functions $\chi'(\omega), \chi''(\omega)$. For the
single-photon Rabi response we get the values which are five times
as little. Typical amplitudes of change of $\alpha_T$ in the
five-photon experiment is almost a factor of 5 as compared to
single-photon measurements giving an estimate for the minimum
decoherence time of the five-photon Rabi oscillations
$\tau_\varphi=0.17\,\mu$s. This confirms the above assumption about
lesser effect of noise in the region of the multi-photon resonances.
In connection with this it should be noted that the measurement of
the charging energy $E_{CP}$ of Hamiltonian (\ref{H-q}) of a
charge-phase qubit in the single-photon resonance mode may bring up
considerably overestimated total capacitance of the island
$C_\Sigma$ because of the charge gate noise.

The characteristic behavior of $\alpha_T(\delta)$ shown in
Figs.\,\ref{fig06} -- \ref{fig08} indicates that if the qubit is
excited at frequencies close to those of the five-photon resonance
then peaks and dips are observed (like in the case of the
single-photon excitation) caused by the alteration of sign of the
Rabi-impedance inserted in the tank circuit when crossing the
resonance region $\hbar\omega_0 \cong\Delta E(\delta,n_g)$. The
increase in the tank circuit quality observed at the single- and
five-photon excitations reflects the fact that the microwave field
energy being transformed by the qubit into Rabi-like oscillations is
transferred to the tank circuit. The decrease in the quality
corresponds to the reverse energy flow.

The interference character of the interaction between the microwave
field with frequency $\omega_0$ and the time-dependent frequency
equivalent energy gap $\omega_g(\delta,n_g)$ of the charge-phase
qubit leads to a sharp dependence of the shape of the signal
characteristics of the qubit detector on the microwave field
parameters. An example of such a dependence which emphasizes high
sensitivity of the multi-photon spectroscopy is shown in
Fig.\,\ref{fig09}. In this case, the peak observed at $\delta=\pi$
and frequency $f=1.435\,$GHz transforms into a dip with shifting the
frequency of the microwave field by $5\,$MHz, i.e. $\Delta
f/f\approx 3.5\!\cdot\!10^{-3}$.

Observed quasiperiodicity of the signal characteristics
$\alpha_T(\delta), V_T(\delta)$ of the charge-phase qubit (see
Fig.\,\ref{fig06}, \ref{fig07}) is determined by the behavior of the
quasienergy (qubit's macrolevels) population in the magnetic and
electromagnetic fields. The rise of $\varepsilon_J(\delta)$ when
deviating the external magnetic flux from $0.5\Phi_0$ leads to
meeting the resonant conditions for resonances of higher order, and
the time-averaged upper level occupation probability reaches its
maximum. If the microwave power is then such that the frequency of
the excited multi-photon Rabi oscillations is close to the tank
circuit frequency, the interaction between the Rabi oscillations
with the tank circuit oscillations will turn again to the parametric
increase (decrease) of the signals. The typical period of such
dependence on the magnetic flux is determined by the product of the
Josephson energies of the qubit's contacts which enter into the
expression for $\varepsilon_J(\delta)$.

\section{Conclusions}
We have described experiments which demonstrated the coherent
Rabi-like oscillations in a charge-phase qubit being under
continuous microwave irradiation involving single- and multi-photon
quantum dynamics in the state of weak coupling to a radio-frequency
tank-detector. Our results show that the charge-phase qubit appears
to be quite sensitive as a quantum parametric detector with
self-pumping provided by inner Rabi generation (Figs.\,2, 6 -- 9).
In the first series of experiments we measured the magnetic flux
dependence of the qubit response, the qubit being irradiated by
microwave field with frequencies in the region of the single-photon
transitions (Fig.\,\ref{fig02}). It was shown by the noise
spectroscopy that the extrema appeared in the signal characteristics
$\alpha_T(\Phi_e/\Phi_0), V_T(\Phi_e/\Phi_0)$ were mainly due to
increase and decrease of the tank circuit quality and reflected the
fact that the power in these (corresponding to the extrema) points
flowed from the microwave field into the tank circuit (peaks) and
vice versa (dips) (Fig.\,\ref{fig03}). With the microwave power at
its optimum value to provide the maximum response, we observed the
resistance Re$Z(\omega)$ inserted from the qubit into the tank
circuit at the extrema points rapidly fell in the region
$T\approx150$ -- 200\,mK approaching to zero in 250 -- 350\,mK
region within the experimental uncertainties (Fig.\,\ref{fig04}). We
believe that the effective noise temperature of the qubit
$T_{eff}\approx 80\,$mK and the decoherence rate in our experiments
are determined by irradiation from the part of the tank circuit and
the transistor placed at $T\approx 1.7\,$K. The development of an
improved type of readout with small backaction effect would be a
major step forward for further improving characteristics of the
qubit detector with self-pumping by intrinsic Rabi generation. At
the same time, the extension of theory to nonzero temperatures would
be of great significance for understanding the temperature effect on
the coherent quantum phenomena in qubits and conditioned by them
detected signals. In second series of the experiments we measured
the qubit responce to variation of the magnetic flux and the
electric charge in microwave field near the five-photon resonance
(Figs.\,\ref{fig06} -- \ref{fig09}). An enhancement of the quality
of the multi-photon resonances, as it was supposed according to the
theory,\cite{Krainov} results in the partial filtration of noise
photons and the increase in amplitude of the signal characteristics
of the detector associated with the multi-photon Rabi oscillations.
The fact that the high transducing steepness is observed at the
point $\delta=\pi$ for some signal characteristics
$\alpha_T(\delta)$ in the region of five-photon resonance points out
to a sharp dependence of quasienergy on microwave power and, hence,
to strong nonlinearity of the system at high powers.\cite{Everitt}

\acknowledgments The authors would like to thank A. Izmalkov, Th.
Wagner, S.N. Shevchenko and O.G. Turutanov for helpful discussions.
The work was partially supported by the Deutsche
Forschungsgemeinschaft under Contract No. KR1172/9-2 and Grant No.
M/189-2007 within the Nanophysics and nanoelectronics" program of
the Ministry of Education and Science of Ukraine.


\newpage
\begin{thebibliography}{99}
\bibitem{R1}
Y. Nakamura, Yu.A. Pashkin, and J.S. Tsai, Phys. Rev. Lett. {\bf
87}, 246601 (2001).

\bibitem{R2}
D. Vion, A. Aassime, A. Cottet, P. Jooyez, H. Pothier, C. Urbina, D.
Esteve, and M.H. Devoret, Science {\bf 296}, 886 (2002).

\bibitem{R3}
I. Chiorescu, Y. Nakamura, C.J.P.M. Harmans, J.E. Mooij, Science
{\bf 299}, 1869 (2003).

\bibitem{R4}
J.M. Martinis, S. Nam, J. Aumentado, and C. Urbina, Phys. Rev. Lett.
{\bf 89}, 117901 (2002).

\bibitem{R5}
S. Saito, M. Thorwart, H. Tanaka, M. Ueda, H. Nakano, K. Semba, and
H. Takayanagi, Phys. Rev. Lett. {\bf 93}, 037001 (2004).

\bibitem{R6}
S. Saito, T. Meno, M. Ueda, H. Tanaka, K. Semba, and H. Takayanagi,
Phys. Rev. Lett. {\bf 96}, 107001 (2006).

\bibitem{R7}
M. Metcalfe, E. Boaknin, V. Manucharyan, R. Vijay, I. Siddiqi, C.
Rigetti, L. Frunzio, R. J. Schoelkopf, and M. H. Devoret, Phys. Rev.
B {\bf 76}, 174516 (2007).

\bibitem{Kanter}
H. Kanter and F.L. Vernon, Jr., J. Appl. Phys. {\bf 43} (7), 3174
(1972).

\bibitem{Likharev}
K.K. Likharev, Dynamics of Josephson Junctions and Circuits, Gordon
$\&$ Breach, Amsterdam (1986).

\bibitem{Smirnov1}
A.Yu. Smirnov, Phys. Rev. B {\bf 68}, 134514 (2003).

\bibitem{Krech1}
W. Krech, D. Born, V. Shnyrkov, Th. Wagner, M. Grajcar, E. Il’ichev,
H.-G. Meyer, and Y. Greenberg, IEEE Trans. on Applied Supercond.
{\bf 15} (2), 876 (2005).

\bibitem{Greenberg}
Ya. S. Greenberg and E. Il'ichev, Phys. Rev. B {\bf 77}, 094513
(2008).

\bibitem{Sisyphus1}
J. Hauss, A. Fedorov, C. Hutter, A. Shnirman, and Gerd Sch\"on,
Phys. Rev. Lett. {\bf 100}, 037003 (2008).

\bibitem{Sisyphus2}
M. Grajcar, S.H.W. van der Ploeg, A. Izmalkov, E. Il'ichev, H.-G.
Meyer, A. Fedorov, A. Shnirman, and Gerd Sch\"on, Nature Physics
{\bf 4}, 612 (2008).

\bibitem{FQ}
A. Izmalkov, S.H.W. van der Ploeg, S.N. Shevchenko, M. Grajcar, E.
Il’ichev, U. H\"ubner, A.N. Omelyanchouk, and H.-G. Meyer, Phys.
Rev. Lett. {\bf 101}, 017003 (2008).

\bibitem{FQ1&2}
S.N. Shevchenko, S.H.W. van der Ploeg, M. Grajcar, E. Il’ichev, A.N.
Omelyanchouk, and H.-G. Meyer, arXiv: 0808.1520v1.

\bibitem{therm}
D. Vion, P.F. Orfila, P. Joyez, D. Esteve, and M.H. Devoret, J.
Appl. Phys. {\bf 77}(6), 2516 (1995).

\bibitem{P-Rabi}
N. Gr{\o}nbech-Jensen and M. Cirillo, Phys. Rev. Lett. {\bf 95},
067001 (2005).

\bibitem{Zorin}
A.B. Zorin, Physica C {\bf 368}, 284 (2002).

\bibitem{Krech2}
W. Krech, M. Grajcar, D. Born, I. Zhilyaev, Th. Wagner, E. Il'chev,
and Ya. Greenberg, Phys. Lett. A {\bf 303}, 352 (2002).

\bibitem{Friedman}
J.R. Friedman and D.V. Averin, Phys. Rev. Lett. {\bf 88}, 050403
(2002).

\bibitem{Zorin2}
A.B. Zorin, Sov. Phys. JETP {\bf 125}(6), 1423 (2004).

\bibitem{IMT1}
R. Rifkin, D.A. Vincent, B.S. Deaver, and P.K. Hansma, J. Appl.
Phys. {\bf 47}, 2645 (1976).

\bibitem{IMT2}
V.I. Shnyrkov, V.A. Khlus, G.M. Tsoi, J. Low Temp. Phys. {\bf
39}(5/6), 477 (1980).

\bibitem{Korotkov}
A.N. Korotkov and D.V. Averin, Phys. Rev. B {\bf 64}, 165310 (2001).

\bibitem{Averin1}
D.V. Averin, in "Exploring the quantum/classical frontier: recent
advance in macroscopic quantum phenomena", Ed. by J.R. Frirdman and
S. Han (Nova Publishes, Hauppauge, NY, 2002), p. 441.

\bibitem{Il'ichev}
E. Il’ichev, N. Oukhanski, A. Izmalkov, Th. Wagner, M. Grajcar,
H.-G. Meyer, A.Yu. Smirnov, A.M. van den Brink, M.H.S. Amin, and
A.M. Zagoskin, Phys. Rev. Lett. {\bf 91}, 097906 (2003).

\bibitem{Sillanpaa}
M. Sillanp\"a\"a, T. Lehtinen, A. Paila, Y. Makhlin, and P. Hakonen,
Phys. Rev. Lett. {\bf 96}, 187002 (2006).

\bibitem{CH}
D. Born, V.I. Shnyrkov, W. Krech, Th. Wagner, E. Il’ichev, M.
Grajcar, U. H\"ubner, and H.-G. Meyer, Phys. Rev. B {\bf 70},
180501(R) (2004).

\bibitem{CH1}
V.I. Shnyrkov, Th. Wagner, D. Born, S.N. Shevchenko, W. Krech, A.N.
Omelyanchouk, E. Il’ichev, and H.-G. Meyer, Phys. Rev. B {\bf 73},
024506 (2006).

\bibitem{LLP}
E.M. Lifshitz, L.D. Landau, Statistical Physics: Course of
Theoretical Physics, Vol. 5 (Butterworth-Heinemann; 3rd edition,
1984), Ch. 12, p. 377.

\bibitem{Everitt}
M.J. Everitt, P. Stiffell, T.D. Clark, A. Vourdas, J.F. Ralph, H.
Prance, and R.J. Prance, Phys. Rev. B {\bf 63}, 144530 (2001).

\bibitem{Krainov}
V.P. Krainov and V.P. Yakovlev, Sov. Phys. JETP {\bf 51}, 1104
(1980).

\bibitem{Makhlin}
Y. Makhlin, G. Sch\"on, and A. Shnirman, Rev. Mod. Phys. {\bf 73},
357 (2001).

\bibitem{Tuominen}
M.T. Tuominen, J.M. Hergenrother, T.S. Tighe, and M. Tinkham, Phys.
Rev. Lett. {\bf 69}, 1997 (1992).

\bibitem{MFLZ}
S.N. Shevchenko, A.S. Kiyko, A.N. Omelyanchouk, and W. Krech, Low
Temp. Phys. {\bf 31}, 569 (2005).

\bibitem{SET-detector}
R.J. Schoelkopf, P. Wahlgren, A.A. Kozhevnikov, P. Delsing, D.E.
Prober, Science {\bf 280}, 1238 (1998).

\bibitem{Flux-detector}
A. Lupascu, S. Saito, T. Picot, P.C. de Groot, C.J.P.M. Harmans and
J.E. Mooij, Nature {\bf 3}, 119 (2007).



\end{thebibliography}
\end{document}